\documentclass[12pt,a4paper]{article}

\usepackage[T1]{fontenc}
\usepackage{lmodern}
\usepackage{amsmath}
\usepackage{amsfonts}
\usepackage{amssymb}
\usepackage{mathrsfs}
\usepackage{physics}
\usepackage{graphicx}
\usepackage{caption}
\usepackage{subcaption}
\usepackage{jheppub}
\usepackage{enumitem}
\usepackage{framed,float}
\usepackage{booktabs}
\usepackage{array}
\usepackage{soul}
\usepackage{multirow}
\usepackage{comment}
\usepackage{braket}
\usepackage{todonotes}
\usepackage{youngtab}
\usepackage{tikz}
\usetikzlibrary{calc,arrows,decorations.markings}

\usepackage{color}

\preprint{}
\title{\boldmath Extended Heun Hierarchy in Quantum Seiberg-Witten Geometry}

\author[a]{Peng Yang,}
\author[a]{Yi-Rong Wang,}
\author[a,b,c,1]{and Kilar Zhang\note{Corresponding Author.}}
\affiliation[a]{Department of Physics and Institute for Quantum Science and Technology, Shanghai University, 99 Shangda Road, Shanghai 200444, China}
\affiliation[b]{Shanghai Key Lab for Astrophysics, Shanghai Normal University, 100 Guilin Road, Shanghai 200234, China}
\affiliation[c]{Shanghai Key Laboratory of High Temperature Superconductors, Shanghai 200444, China}
\emailAdd{yangpeng522@shu.edu.cn}
\emailAdd{wyr2024@shu.edu.cn}
\emailAdd{kilar@shu.edu.cn}

\abstract{
We investigate the quantum geometry of the Seiberg-Witten curve for $\mathcal{N}=2$, $\mathrm{SU(2)}^n$ linear quiver gauge theories. 
By applying the Weyl quantization prescription to the algebraic curve, we derive the corresponding second-order differential equation and demonstrate that it is isomorphic to the Extended Heun Equation with $n+3$ regular singular points. 
The physical parameters of the gauge theory are linked to the canonical coefficients of the Heun equation via a polynomial representation of the Seiberg-Witten curve.
This framework provides the necessary mathematical foundation to apply non-perturbative gauge-theoretic techniques, such as instanton counting, 
to spectral problems in gravitational physics, most notably for higher-dimensional black holes.
}

\begin{document}

\maketitle
\flushbottom

\setcounter{footnote}{0}

\section{Introduction}
\label{sec:intro}

The exact solution of $\mathcal{N}=2$ supersymmetric gauge theories in four dimensions constitutes a cornerstone of modern theoretical physics. Since the seminal work of Seiberg and Witten \cite{Seiberg:1994rs, Seiberg:1994aj}, it has been understood that the low-energy effective dynamics of these theories is encoded in the complex geometry of an auxiliary Riemann surface $\Sigma$, the Seiberg-Witten (SW) curve, and a specific meromorphic differential $\lambda_{SW}$. While the classical geometry determines the prepotential of the theory via period integrals, recent developments have established a profound connection between the quantization of these algebraic curves and the spectral theory of second-order differential equations \cite{Aminov:2020yma,Aminov:2023jve,Nekrasov:2009rc}.

A particularly rigorous formulation of these geometries was introduced by Gaiotto \cite{Gaiotto:2009we}, who reconstructed a large class of $\mathcal{N}=2$ superconformal field theories by compactifying the $(2,0)$ six-dimensional theory on a Riemann surface with punctures. In this framework, the linear quiver gauge theories, consisting of a chain of $n$ $\mathrm{SU}(2)$ gauge groups with bifundamental matter, are realized on a sphere with $n+3$ regular singularities \cite{Witten:1997sc,Gaiotto:2009we,Alday:2009aq}. The physical data of the theory, such as coupling constants and mass parameters, are elegantly encoded in the complex structure moduli of the punctured sphere and the residues of the differential at the singularities.

Recently, this geometric structure has found unexpected and powerful applications in gravitational physics. It has been proposed that the master wave equations governing the linear perturbations of black holes are isomorphic to the quantum SW curves of specific supersymmetric gauge theories \cite{Aminov:2020yma,Bianchi:2021mft,Consoli:2022eey,Gorsky:2015toa,CarneirodaCunha:2015qln}. In this correspondence, the quasinormal mode (QNM) \cite{Kokkotas:1999bd,Berti:2009kk,Konoplya:2011qq,Horowitz:1999jd} frequencies of the black hole are mapped to the spectral parameters of the quantum curve. This duality transforms the problem of computing the gravitational spectrum into the problem of solving exact quantization conditions in the gauge theory. Specifically, for theories with known duals, the spectral problem can be solved analytically using the Nekrasov-Shatashvili (NS) limit of the instanton partition function \cite{Nekrasov:2009rc,Nekrasov:2002qd}, where the quantized periods of the curve are identified with the derivatives of the NS free energy \cite{Gukov:2011qp,Mironov:2009uv,Nekrasov:2009rc}.

However, the existing literature on this correspondence has largely focused on systems governed by the standard Heun equation, corresponding to $N_f=4$ Supersymmetric Quantum Chromodynamics (SQCD), or its confluent limits ($N_f \le 3$) \cite{Batic:2007it,Alday:2009aq,Staicova:2016awo,Aminov:2020yma,Lei:2023mqx,Ge:2024jdx, Ge:2025yqk}. A significant gap remains regarding gravitational backgrounds with richer singularity structures. Notably, the wave equations for higher-dimensional Schwarzschild-(A)dS black holes in $d$ dimensions ($d > 4$) possess $d+1$ regular singularities \cite{Kodama:2003jz}, exceeding the four singularities of the standard Heun equation. Mathematically, these systems require the machinery of what we term the Extended Heun Equation (EHE) \cite{Slavyanov:2000sf}. In the language of gauge theory, the natural dual candidates are the linear quiver theories of rank $n=d-2$, which possess exactly $n+3$ regular singularities \cite{Gaiotto:2009we}.

In this paper, we bridge this gap by performing a systematic quantization of the SW geometry for generic $\mathrm{SU(2)}$ linear quiver gauge theories. We adopt the polynomial formulation of the curve, where the physical parameters, masses, Coulomb branch moduli, and couplings, are embedded in the coefficients of defining polynomials. By applying the Weyl quantization prescription directly to the algebraic curve equation, we derive the corresponding second-order Schrödinger-like differential equation.

Our main result is the establishment of a precise mapping that translates the polynomial data of the SW curve into the canonical coefficients of EHE. We demonstrate that the quantum geometry of the $\mathrm{SU}(2)^n$ linear quiver is isomorphic to a EHE with $n+3$ singularities. Crucially, we derive explicit formulas for the accessory parameters of the differential equation in terms of the polynomial coefficients. Since these polynomials are directly constructed from the gauge theory parameters 
, this result completes the dictionary between the physics of the linear quiver and the spectral data of the EHE.

While the primary focus of this work is the structural derivation of the quantum differential equation, this formalism provides the necessary foundation for extending the analytic computation of black hole spectra to higher dimensions. By identifying the EHE derived here with the master equation of a $d$-dimensional black hole, one can utilize the quiver gauge theory description to solve the spectral problem. 
In this context, the QNM quantization corresponds to the exact Bohr-Sommerfeld quantization of the curve periods \cite{Mironov:2009uv}, identified with the derivatives of the NS free energy \cite{Nekrasov:2009rc}.
Our derivation ensures that the parameters entering these free energy calculations are correctly identified with the geometric parameters of the black hole.

The paper is organized as follows. In Section \ref{sec:classical_geometry}, we review the classical geometry of linear quivers, detailing how the physical parameters are encoded in the polynomial coefficients of the SW curve. Section \ref{sec:quantum_geometry} presents the core of our analysis: we apply the Weyl quantization to transform the algebraic curve into a second-order differential equation, analyze its singularity structure via partial fraction decomposition, and prove its isomorphism to EHE. We also provide a rational polynomial representation of the potential that is well-suited for practical computations. Finally, in Section \ref{sec:implications}, we discuss the physical implications of our results, specifically outlining the dictionary between higher-dimensional black holes and linear quivers, and describing how the NS limit can be employed to determine the QNM spectrum. Technical details regarding the derivation of the operator and residue calculations are provided in the Appendices.

\section{Classical Seiberg-Witten Geometry for Linear Quivers}
\label{sec:classical_geometry}

In this section, we review the classical geometry of the SW curve for $\mathcal{N}=2$ linear quiver gauge theories. We adopt the formulation introduced by Gaiotto \cite{Gaiotto:2009we}, which encodes the physical data, coupling constants, mass parameters, and Coulomb branch moduli, into the coefficients of a polynomial equation. This setup provides the classical foundation for the quantization procedure discussed in Section \ref{sec:quantum_geometry}.

We study the $A_1$ class $\mathcal{S}$ theory associated with a sphere with punctures, which admits a linear quiver description with $n$ $\mathrm{SU}(2)$ gauge nodes.
The SW curve $\Sigma$ is realized as a ramified covering over the Riemann sphere (parametrized by $y$) and is defined by the algebraic equation:
\begin{equation}
    \label{eq:SW_curve_poly}
    F_2(y)x^2 - F_1(y)x - F_0(y) = 0 \,.
\end{equation}
Here, $x$ is the SW differential coordinate, related to the canonical one-form by $\lambda_{SW} = x dy/y$. The coefficients $F_i(y)$ are polynomials in $y$ of degree $n+1$:
\begin{equation}
\begin{aligned}
    F_2(y) &= \prod_{k=0}^{n} (y-y_k) \quad (\text{with } y_0 \equiv 1) \,, \\
    F_1(y) &= \sum_{i=0}^{n+1} c_i y^i \,, \\
    F_0(y) &= \sum_{i=0}^{n+1} d_i y^i \,.
\end{aligned}
\end{equation}
The roots of $F_2(y)$, denoted by $\{y_k\}_{k=0}^n$, together with the points at $0$ and $\infty$, define the $n+3$ singular points (punctures) on the base Riemann sphere. The complex structure moduli of this punctured sphere encode the UV gauge couplings $\tau_i$ via the cross-ratios of these points.

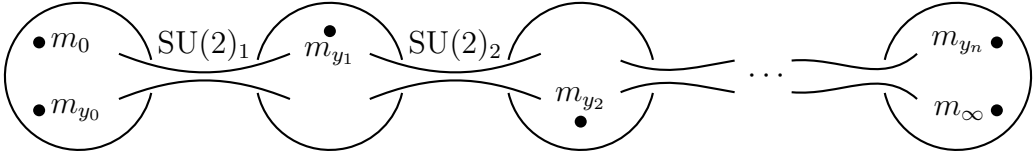
\begin{figure}[htbp]
	\centering
	\begin{tikzpicture}[thick, scale=1.5, >=stealth]
		\def\R{0.65}\def\dist{2.2}\def\gap{13}\def\tubeStart{0.35}\def\tubeY{0.20}\def\bend{22}
		\coordinate (C1) at (0,0);\coordinate (C2) at (\dist,0);\coordinate (C3) at (2*\dist,0); \coordinate (C4) at (3.5*\dist,0); \coordinate (MidEllipsis) at (2.75*\dist,0);
		\draw (C1) +(\gap:\R) arc (\gap:360-\gap:\R);
		\fill ($(C1)+(-0.35, 0.3)$) circle (1.5pt) node[right] {$m_0$};
		\fill ($(C1)+(-0.35, -0.3)$) circle (1.5pt) node[right] {$m_{y_0}$};
		\draw (C2) +(\gap:\R) arc (\gap:180-\gap:\R);
		\draw (C2) +(180+\gap:\R) arc (180+\gap:360-\gap:\R);
		\fill ($(C2)+(0, 0.4)$) circle (1.5pt) node[below] {$m_{y_1}$};
		\draw (C3) +(\gap:\R) arc (\gap:180-\gap:\R);
		\draw (C3) +(180+\gap:\R) arc (180+\gap:360-\gap:\R);
		\fill ($(C3)+(0, -0.4)$) circle (1.5pt) node[above] {$m_{y_2}$};
		\draw (C4) +(180+\gap:\R) arc (180+\gap:540-\gap:\R);
		\fill ($(C4)+(0.35, 0.3)$) circle (1.5pt) node[left] {$m_{y_n}$};
		\fill ($(C4)+(0.35, -0.3)$) circle (1.5pt) node[left] {$m_\infty$};
		\draw ($(C1)+(\tubeStart, \tubeY)$)
		to[out=-\bend, in=180+\bend]
		node[midway, above] {$\mathrm{SU(2)_1}$}
		($(C2)+(-\tubeStart, \tubeY)$);
		\draw ($(C1)+(\tubeStart, -\tubeY)$)
		to[out=\bend, in=180-\bend]
		($(C2)+(-\tubeStart, -\tubeY)$);
		\draw ($(C2)+(\tubeStart, \tubeY)$)
		to[out=-\bend, in=180+\bend]
		node[midway, above] {$\mathrm{SU(2)_2}$}
		($(C3)+(-\tubeStart, \tubeY)$);
		\draw ($(C2)+(\tubeStart, -\tubeY)$)
		to[out=\bend, in=180-\bend]
		($(C3)+(-\tubeStart, -\tubeY)$);
		\draw ($(C3)+(\tubeStart, \tubeY*0.7)$) to[out=-\bend, in=190] ($(MidEllipsis)+(-0.3, \tubeY*0.7)$);
		\draw ($(C3)+(\tubeStart, -\tubeY*0.7)$) to[out=\bend, in=170] ($(MidEllipsis)+(-0.3, -\tubeY*0.7)$);
		\node at (MidEllipsis) {$\cdots$};
		\draw ($(MidEllipsis)+(0.2, \tubeY*0.7)$) to[out=5, in=200+\bend] ($(C4)+(-\tubeStart, \tubeY)$);
		\draw ($(MidEllipsis)+(0.2, -\tubeY*0.7)$) to[out=-5, in=160-\bend] ($(C4)+(-\tubeStart, -\tubeY)$);
	\end{tikzpicture}
	
	\caption{The Riemann surface for the $\mathrm{SU}(2)^n$ linear quiver viewed as a gluing of $n+1$ three-punctured spheres. The tubes represent the gauge nodes $\mathrm{SU}(2)_k$, and the marked points correspond to the mass singularities $m_i$ discussed in the text.}
	\label{fig:gaiotto_curve_decomposition}
\end{figure}

The physical mass parameters are determined by the behavior of the SW differential $\lambda_{SW}$ at these singular points. In the standard frame of the linear quiver, the coordinate singularities at the origin ($y=0$) and infinity ($y=\infty$) correspond to the fundamental hypermultiplets attached to the far ends of the quiver chain. Their masses are determined by the asymptotic difference of the roots $x_{\pm}(y)$ of the curve equation. Specifically, at $y=0$, the mass $m_0$ is given by:
\begin{equation}
    m_0 = x_+(0) - x_-(0) = \frac{\sqrt{c_0^2 + 4 F_2(0) d_0}}{|F_2(0)|} \,,
\end{equation}
where $F_2(0) = (-1)^{n+1} \prod_{k=0}^n y_k$. Similarly, at $y \to \infty$, the mass $m_\infty$ is:
\begin{equation}
    m_\infty = \lim_{y\to\infty} (x_+(y) - x_-(y)) = \sqrt{c_{n+1}^2 + 4 d_{n+1}} \,.
\end{equation}

The remaining mass parameters are encoded in the roots $\{y_k\}$ of $F_2(y)$, where the differential exhibits simple poles. 
Physically, these singularities encode the masses of the bifundamental hypermultiplets connecting the gauge nodes, as well as the remaining fundamental masses. 
In the vicinity of the root $y = y_k$, the differential diverges, and the corresponding mass parameter is defined by the residue:
\begin{equation}
    m_{y_k} = \frac{1}{2\pi i} \oint_{y_k} x \frac{dy}{y} = \frac{F_1(y_k)}{y_k F_2'(y_k)} = \frac{\sum_{i=0}^{n+1} c_i y_k^i}{y_k \prod_{j \neq k} (y_k - y_j)} \,.
\end{equation}
This formula applies to all $k=0, \dots, n$, thereby explicitly mapping the polynomial coefficients to the full set of $n+3$ physical mass parameters.

While the mass parameters are fixed constants of the theory, the Coulomb branch moduli $\boldsymbol{u} = \{u_1, \dots, u_n\}$ represent the vacuum expectation values of the scalar fields. To disentangle the moduli from the masses, it is convenient to transform the curve into a standard form by completing the square.

Defining a shifted coordinate $\tilde{x} = x - \frac{F_1(y)}{2F_2(y)}$, the curve equation \eqref{eq:SW_curve_poly} transforms into:
\begin{equation}
    \tilde{x}^2 = \phi_2(y) \equiv \frac{P_{2n+2}(y)}{4 F_2(y)^2} \,,
\end{equation}
where the numerator polynomial is the discriminant of the original equation:
\begin{equation}
    P_{2n+2}(y) = F_1(y)^2 + 4 F_2(y) F_0(y) \,.
\end{equation}
In this quadratic differential form $\phi_2(y)\,dy^2$, the mass parameters are fixed by the coefficients of the double poles at the singularities. Any deformation of the polynomial $P_{2n+2}(y)$ that preserves these double pole coefficients corresponds to a variation along the Coulomb branch.

Specifically, we can decompose the polynomial $F_0(y)$ to isolate the moduli dependence. By extracting the leading and trailing terms which contribute to $m_\infty$ and $m_0$ respectively, we write:
\begin{equation}
    F_0(y) = d_{n+1} y^{n+1} + y U_{n-1}(y) + d_0 \,,
\end{equation}
where $U_{n-1}(y) = \sum_{j=1}^{n} d_j y^{j-1}$. 
Substituting this decomposition into $P_{2n+2}(y)$, we find that the intermediate coefficients $\{d_1, \dots, d_n\}$ alter neither the residues at the punctures $y_k$ (which are fixed by $F_1^2$) nor the asymptotic mass differences. Consequently, we identify these intermediate coefficients as the algebraic coordinates of the Coulomb branch moduli: 
\begin{equation}
    u_k = d_k \quad \text{for } k=1, \dots, n \,.
\end{equation}
This precise identification allows us to map the accessory parameters of the EHE derived in later sections directly to the physical moduli space of the gauge theory.

\section{Quantum Geometry and the Extended Heun Equation}
\label{sec:quantum_geometry}

In this section, we promote the classical SW geometry described in Section \ref{sec:classical_geometry} to a quantum mechanical system. We apply the Weyl quantization prescription to the algebraic curve, converting it into a second-order differential operator. Through a standard Liouville transformation, we map this operator to a canonical Schrödinger equation and identify the effective potential. We then demonstrate that for an $\mathrm{SU}(2)^n$ linear quiver, this potential corresponds precisely to the EHE with $n+3$ regular singular points, and we derive the explicit mapping between the gauge theory parameters and the canonical Heun coefficients.

\subsection{Weyl Quantization and the Schrödinger Form}

We start with the classical curve equation \eqref{eq:SW_curve_poly}, which we rewrite using the explicit polynomial coefficients. We denote the coefficients of the leading polynomial $F_2(y)$ as $F_2(y) = \sum_{k=0}^{n+1} f_k y^k$. 
The quantization is performed by promoting the coordinates $(y, x)$ to non-commutative operators satisfying $[\hat{x}, \hat{y}] = \hbar \hat{y}$. In the coordinate representation where $\hat{y}$ acts as multiplication by $y$, the momentum operator is realized as $\hat{x} = \hbar y \partial_y$.

We adopt the Weyl ordering prescription, which ensures Hermiticity and maps classical monomials $y^k x^m$ to symmetric operators. For the quadratic form of the SW curve, this is equivalent to the specific shifts:
\begin{equation}
    y^k x \to y^k \left(\hbar y \partial_y + \frac{k\hbar}{2}\right) \,, \quad y^k x^2 \to y^k \left( \hbar^2 y^2 \partial_y^2 + \hbar^2(1+k) y \partial_y + \frac{\hbar^2 k^2}{4} \right) \,.
\end{equation}
Implementing this quantization procedure on the classical constraint \eqref{eq:SW_curve_poly} yields a second-order linear differential equation for the wave function $\Psi(y)$. 
By expanding the momentum operator $\hat{x}$ and collecting terms by derivative order, we obtain a linear differential equation of the form:
\begin{equation}
    \label{eq:general_ODE_operator}
    \left[ A(y) \frac{d^2}{dy^2} + B(y) \frac{d}{dy} + C(y) \right] \Psi(y) = 0 \,.
\end{equation}
The coefficient functions $A(y)$, $B(y)$, and $C(y)$ are completely determined by the geometry of the curve (see Appendix \ref{app:derivation} for the derivation):
\begin{subequations}
\begin{align}
    A(y) &= \hbar^2 y^2 F_2(y) \,, \\
    B(y) &= \hbar^2 y \left( F_2(y) + y F_2'(y) \right) - \hbar y F_1(y) \,, \\
    C(y) &= \frac{\hbar^2}{4} y (y F_2'(y))' - \frac{\hbar}{2} y F_1'(y) - F_0(y) \,.
\end{align}
\end{subequations}

To extract the physical content of the differential equation and facilitate the singularity analysis, it is advantageous to eliminate the first-derivative term. We introduce the rescaled wave function $\psi(y)$ via the transformation $\Psi(y) = N(y) \psi(y)$. The condition that the first derivative vanishes fixes the scaling factor $N(y)$ to be:
\begin{equation}
    N(y) = \exp\left( - \int^y \frac{B(z)}{2A(z)} \, dz \right) = \frac{1}{\sqrt{y F_2(y)}} \exp\left( \frac{1}{2\hbar} \int^y \frac{F_1(z)}{z F_2(z)} \, dz \right) \,.
\end{equation}
Under this transformation, the operator equation \eqref{eq:general_ODE_operator} reduces to the standard one-dimensional Schrödinger form:
\begin{equation}
    \label{eq:standard_schrodinger}
    \frac{d^2\psi}{dy^2} + V(y)\psi(y) = 0 \,.
\end{equation}
The effective quantum potential $V(y)$ is given by the invariant of the differential equation:
\begin{equation}
    V(y) = \frac{C(y)}{A(y)} - \frac{1}{4}\left(\frac{B(y)}{A(y)}\right)^2 - \frac{1}{2} \frac{d}{dy}\left(\frac{B(y)}{A(y)}\right) \,.
\end{equation}
Substituting the explicit expressions for $A, B, C$ into this invariant, we derive the potential in terms of the gauge theory polynomials:
\begin{equation}
    \label{eq:potential_explicit}
    V(y) = -\frac{F_0(y)}{\hbar^2 y^2 F_2(y)} - \frac{1}{4\hbar^2 y^2} \left( \frac{F_1(y)}{F_2(y)} \right)^2 + \left[ \frac{1}{4y^2} - \frac{F_2''(y)}{4F_2(y)} - \frac{F_2'(y)}{4yF_2(y)} + \frac{(F_2'(y))^2}{4F_2(y)^2} \right] \,.
\end{equation}
This explicit expansion isolates the classical singularity of order $\hbar^{-2}$ from the subleading quantum corrections (in brackets). The resulting potential $V(y)$ provides a complete description of the quantum SW curve in the Schrödinger frame.

\subsection{Analytic Structure and Rational Representation}

Having established the Schrödinger form, we now analyze its analytic structure. The $\mathrm{SU}(2)^n$ linear quiver gauge theory is characterized by a SW curve with singular points at $\{0, 1, y_1, \dots, y_n, \infty\}$. Consequently, the effective potential $V(y)$ derived above exhibits a specific pole structure on the Riemann sphere. To make the connection to the standard literature on Heun functions explicit \cite{Slavyanov:2000sf}, we adopt the partial fraction decomposition for a EHE with $N=n+3$ singularities.

We decompose the potential into characteristic quadratic poles (determining the exponents) and simple poles (associated with accessory parameters). Imposing the standard constraint that cross-terms between non-origin singularities vanish, the canonical form of the potential is:
\begin{equation}
    \label{eq:EHE_canonical}
    V(y) = \frac{S_0}{y^2} + \sum_{k=0}^{n} \left( \frac{S_{y_k}}{(y-y_k)^2} + \frac{K_{k}}{y(y-y_k)} \right) \,.
\end{equation}
Here, the set of singularities is denoted by $\{y_k\}_{k=0}^n$ with the convention $y_0 \equiv 1$. 

The strength of the singularities is governed by the coefficients of the quadratic poles, $S_{y_k}$ and $S_0$. By comparing the leading-order behavior of our gauge-theoretic potential \eqref{eq:potential_explicit} near each singularity $y \to y_k$ and $y \to 0$, we find that these coefficients are completely determined by the values of the polynomials and their derivatives at the singular points:
\begin{equation}
    S_{y_k} = \frac{1}{2}\Lambda_{y_k} - \frac{1}{4}\Lambda_{y_k}^2 \,, \qquad 
    S_0 = -\frac{F_0(0)}{\hbar^2 F_2(0)} + \frac{1}{2}\Lambda_0 - \frac{1}{4}\Lambda_0^2 \,,
\end{equation}
where we have defined the dimensionless shift parameters $\Lambda$:
\begin{equation}
    \Lambda_{y_k} \equiv 1 - \frac{F_1(y_k)}{\hbar y_k F_2'(y_k)} \,, \qquad 
    \Lambda_0 \equiv 1 - \frac{F_1(0)}{\hbar F_2(0)} \,.
\end{equation}
These expressions generalize the relation between pole coefficients and physical mass parameters, previously established for the single gauge node case ($n=1$), to linear quivers of arbitrary rank. 
The explicit residue calculations leading to these pole coefficients are detailed in Appendix \ref{app:residue_calc}.

The coefficients of the simple poles, $K_k$, are identified as the accessory parameters. Unlike the quadratic pole coefficients, these are not fixed solely by local kinematic data but are related to the global properties of the solution, encoding dynamical data such as the Coulomb branch moduli or spectral eigenvalues. 
They satisfy the sum rule $\sum K_k = 0$, ensuring the potential decays appropriately at infinity. Explicitly, they are defined by the residue $K_k = y_k \, \text{Res}_{y=y_k} \, V(y)$. Using the invariant form of the potential expressed in terms of $A, B$, and $C$, we can derive a direct formula for computing these parameters:
\begin{equation}
    K_k = y_k \left. \left[ \frac{C(y)}{A'(y)} - \frac{B(y)}{2A'(y)}\left( \frac{B'(y)}{A'(y)} - \frac{B(y)A''(y)}{2(A'(y))^2} \right) \right] \right|_{y=y_k} \,.
\end{equation}
This formula allows for the computation of the accessory parameters directly from the polynomial data of the SW curve.

To explicitly validate the identification of this system as an EHE, we adopt the standard form for a Fuchsian equation with $N=n+3$ singularities. Let the set of finite singularities be $\Xi = \{0, y_0, y_1, \dots, y_n\}$, which we denote generally by $\xi_j$ (where $j=0, \dots, n+1$). The EHE governing a function $H(y)$ reads:
\begin{equation}
    \label{eq:EHE_standard_form}
    \frac{d^2 H}{dy^2} + \left( \sum_{j=0}^{n+1} \frac{\gamma_j}{y - \xi_j} \right) \frac{d H}{dy} + \frac{\mathcal{Q}(y)}{\prod_{j=0}^{n+1} (y - \xi_j)} H(y) = 0 \,.
\end{equation}
Here, $\mathcal{Q}(y) = \alpha \beta y^{n} + \sum_{k=0}^{n-1} q_k y^k $, which ensures the regularity of the solution at infinity. The parameters $\gamma_j$ determine the characteristic exponents at finite singularities, subject to the constraint $\sum_{j} \gamma_j = \alpha + \beta + 1$.

To map this to our quantum curve, we must eliminate the first derivative term $\frac{dH}{dy}$. We perform the transformation $H(y) = \Omega(y)^{-1} \psi(y)$, where the scaling factor $\Omega(y)$ is determined by:
\begin{equation}
    \Omega(y) = \prod_{j=0}^{n+1} (y - \xi_j)^{\gamma_j/2} \,.
\end{equation}
Under this transformation, equation \eqref{eq:EHE_standard_form} becomes the Schrödinger-like equation $\psi''(y) + Q_{\text{EHE}}(y)\psi(y) = 0$. The invariant potential $Q_{\text{EHE}}(y)$ is explicitly given by:
\begin{equation}
    \label{eq:EHE_potential_invariant}
    Q_{\text{EHE}}(y) = \sum_{j=0}^{n+1} \left[ \frac{\frac{1}{4}(1 - \theta_j^2)}{(y - \xi_j)^2} + \frac{C_j}{y - \xi_j} \right] \,,
\end{equation}
where $\theta_j \equiv 1 - \gamma_j$ represents the exponent difference at each singularity. The coefficients $C_j$ are determined by:
\begin{equation}
    C_j = \frac{\mathcal{Q}(\xi_j)}{\prod_{k \neq j} (\xi_j - \xi_k)} - \frac{\gamma_j}{2} \sum_{k \neq j} \frac{\gamma_k}{\xi_j - \xi_k} \,.
\end{equation}

We can now formulate an exact mapping between the gauge theory and the canonical Heun formalism by comparing $Q_{\text{EHE}}(y)$ with the derived potential $V(y)$. To make the comparison manifest, we expand the partial fraction form of $V(y)$ in equation \eqref{eq:EHE_canonical}:
\begin{equation}
    V(y) = \frac{S_0}{y^2} - \sum_{k=0}^n \frac{K_k}{y_k y} + \sum_{k=0}^{n} \left[ \frac{S_{y_k}}{(y-y_k)^2} + \frac{K_k/y_k}{y-y_k} \right] \,.
\end{equation}
The mapping is established by matching the coefficients of the quadratic and simple poles for each singularity type:
\begin{equation}
    \theta_{y_k} = |1 - \Lambda_{y_k}|\,,\quad C_{y_k} = \frac{K_k}{y_k} \,.
\end{equation}
At the origin:
\begin{equation}
    \theta_0 = \sqrt{1 - 4S_0} \,,\quad C_0 = - \sum_{k=0}^n \frac{K_k}{y_k} \,.
\end{equation}
This identification satisfies the Fuchsian constraint $\sum_{j} C_j = 0$ automatically.

Thus, we have proved that the quantum geometry of the $\mathrm{SU}(2)^n$ linear quiver is isomorphic to the EHE.

While the partial fraction form \eqref{eq:EHE_canonical} is ideal for analyzing monodromies, practical computations, such as instanton counting \cite{Nekrasov:2002qd} or numerical evaluation of QNMs, often benefit from a single rational function representation. The potential can be recast as:
\begin{equation}
    \label{eq:V_rational}
    V(y) = \frac{\mathcal{P}(y)}{4 \hbar^2 y^2 [F_2(y)]^2} \,,
\end{equation}
where the numerator $\mathcal{P}(y)$ is a polynomial of degree $2n+2$ that encapsulates the entire quantum geometry of the system. Substituting the explicit definitions of the coefficient functions $A(y)$, $B(y)$, and $C(y)$ into the invariant form, and performing a systematic expansion in powers of $\hbar$, we derive a compact algorithm for generating $\mathcal{P}(y)$. If we write $\mathcal{P}(y) = \sum_{m} \mathcal{A}_m y^m$, the coefficients $\mathcal{A}_m$ are given by a convolution of the gauge theory parameters (see Appendix \ref{app:polynomial_derivation} for the derivation):
\begin{equation}
    \label{eq:Ak_coefficients}
    \mathcal{A}_m = \hbar^2 \sum_{j=0}^{m} f_j f_{m-j} \left[ 1 - \frac{1}{2}(2j - m)^2 \right] - \sum_{j=0}^{m} \left( c_j c_{m-j} + 4 d_j f_{m-j} \right) \,,
\end{equation}
where $f_j, c_j, d_j$ are the coefficients of the polynomials $F_2, F_1, F_0$ respectively. This algebraic structure provides a direct map from the gauge theory data (masses, couplings, and Coulomb moduli encoded in $F_i$) to the EHE.

\section{Implications for Black Hole Spectroscopy}
\label{sec:implications}

The extraction of the EHE configuration from the SW curves of the $\mathrm{SU}(2)^n$ linear quiver gauge theory is not merely a formal mathematical result. It establishes a precise dictionary between the non-perturbative dynamics of supersymmetric gauge theories and the spectral problems of gravitational physics. 
In this section, we outline a specific physical scenario where our result, particularly the extension to $n+3$ singularities, is essential, and sketch the analytic method for computing the spectrum.

\subsection{The Gauge/Gravity Dictionary in Higher Dimensions}

While the reduction of the four-dimensional Schwarzschild-(A)dS black hole master equation to the standard Heun equation, corresponding to $N_f=4$ SQCD, has been identified based on the singularity structure \cite{Aminov:2020yma}, the extension to higher dimensions introduces a complexity that transcends the classical Heun framework. In this context, the EHE derived in Section \ref{sec:quantum_geometry} serves as the essential bridge, explicitly realizing the correspondence predicted in the literature.

We study vector-type gravitational perturbations of Schwarzschild-(A)dS black holes in $d$ dimensions ($d > 4$), employing the Kodama-Ishibashi formalism \cite{Kodama:2003jz}. The radial dynamics are governed by a master equation determined by the lapse function $f(r) = 1 - (r_0/r)^{d-3} - \Lambda r^2$. A crucial observation, noted in earlier works on the subject \cite{Aminov:2020yma, Gaiotto:2009we}, is that the singularity structure of this equation dictates the matter content of the dual gauge theory. For a non-vanishing cosmological constant, the master equation possesses regular singularities at the origin $r=0$, at spatial infinity $r=\infty$, and at the $d-1$ roots of $f(r)=0$. This results in a Fuchsian differential equation with a total of $d+1$ regular singular points.

As established in the previous section, the EHE associated with an $\mathrm{SU}(2)^n$ linear quiver characterizes a system with $n+3$ regular singularities. By identifying the rank of the quiver hierarchy with the spacetime dimension as $n = d-2$, we obtain a precise match between the $d+1$ singularities of the black hole master equation and the fundamental structure of the $\mathrm{SU(2)}^{d-2}$ quiver theory. This confirms that the spectral problem of a $d$-dimensional black hole is encoded in the quantum geometry of a linear quiver with $d-2$ gauge nodes.

Furthermore, based on the discussion in Section \ref{sec:classical_geometry}, the polynomial coefficients of the SW curve naturally encode the physical parameters. Since both the high-dimensional black hole wave equation and the quantum SW curve can be transformed into the canonical EHE form, comparing their coefficients establishes a precise dictionary. Through this map, the geometric quantities, such as the mass parameter $r_0$, cosmological constant $\Lambda$, and angular momentum, are translated into the gauge couplings $\mathbf{q}$, hypermultiplet masses $\mathbf{m}$, and Coulomb branch moduli $\mathbf{u}$. Consequently, the determination of the QNM frequencies $\omega$ is mapped to the spectral problem of the quantum curve. This correspondence allows one to impose the exact quantization condition \cite{Imaizumi:2022dgj,Hatsuda:2019eoj}, constructed analytically using the NS free energy and instanton counting \cite{Nekrasov:2009rc}, to derive the defining equations for the frequency spectrum.

\subsection{Quantization Condition via the Nekrasov-Shatashvili Limit}
\label{sec:quantization_condition}

The isomorphism between the EHE governing the black hole perturbations and the quantum SW curve for $\mathrm{SU(2)}^n$ linear quiver theories provides a rigorous analytic framework for determining the QNM spectrum. Unlike the simple $\mathrm{SU(2)}$ case considered in previous literature \cite{Hatsuda:2019eoj} where a single Coulomb modulus dictates the dynamics, the higher-dimensional black hole backgrounds correspond to quiver theories with $n$ gauge nodes, characterized by a vector of Coulomb branch moduli $\mathbf{a} = (a_1, \dots, a_n)$. Consequently, the condition for the existence of QNMs is elevated from a single scalar constraint to a system of coupled quantization conditions on the gauge theory side.

This framework is formulated within the NS limit of the $\Omega$-background, where $\epsilon_1 = \hbar$ and $\epsilon_2 \to 0$. The non-perturbative contribution to the physics is encoded in the NS free energy $\mathcal{F}^{\text{NS}}(\mathbf{a}, \mathbf{m}, \mathbf{q}, \hbar)$, defined by the limit $\mathcal{F}^{\text{NS}} = \lim_{\epsilon_2 \to 0} \epsilon_2 \ln \mathcal{Z}_{\text{inst}}$ \cite{Nekrasov:2009rc}. The condition for constructive interference of the wave function along the non-trivial cycles of the SW curve imposes that the dual periods are quantized. For an $n$-node quiver, this yields a set of $n$ coupled equations determining the physical moduli:
\begin{equation}
    \label{eq:quantization_condition_vector}
     \frac{\partial \mathcal{F}^{\text{NS}}(\mathbf{a}, \mathbf{m}, \mathbf{q})}{\partial a_i} = 2\pi \hbar \left( n_i + \frac{1}{2} \right) \,, \quad i = 1, \dots, n \,,
\end{equation}
where $\mathbf{n} = (n_1, \dots, n_n) \in (\mathbb{Z}_{\ge 0})^n$ is a vector of integers labeling the overtone numbers associated with each gauge node. The physical interpretation of these integers depends on the specific topology of the potential barriers; typically, one determines the radial overtone of the black hole mode, while the others may be constrained by hidden symmetries or separation constants in the gravity dual.

To utilize Eq.~\eqref{eq:quantization_condition_vector} for computing the eigenfrequencies $\omega$, one must bridge the gap between the gauge theory moduli $\mathbf{a}$ and the black hole parameters. On the gravity side, the master wave equation, when cast into the extended Heun form, allows us to extract $n$ accessory or energy-like parameters, denoted as $\mathbf{u}^{\text{BH}} = (u_1^{\text{BH}}, \dots, u_n^{\text{BH}})$. These parameters are explicit functions of the frequency $\omega$ and the black hole geometry, $u_i^{\text{BH}} = u_i^{\text{BH}}(\omega)$. 
On the gauge theory side, the corresponding expectation values $\mathbf{u}^{\text{gauge}}$ (classically related to quadratic Casimirs $\langle \text{tr} \phi_i^2 \rangle$) are linked to the Coulomb moduli $\mathbf{a}$ through the quantum mirror map \cite{Aganagic:2011mi,Grassi:2019coc}. This connection is established via the generalized Matone relations \cite{Matone:1995rx,Flume:2004rp}, which identify the order parameters $\mathbf{u}^{\text{gauge}}$ with the logarithmic derivatives of the instanton free energy with respect to the counting parameters $\mathbf{q}$. Consequently, using the explicit expansion of $\mathcal{F}^{\text{NS}}$ derived from instanton counting, one obtains $\mathbf{u}^{\text{gauge}}$ as a series in $\mathbf{q}$:
\begin{equation}
    \label{eq:quantum_mirror_map}
    u_i^{\text{gauge}} = a_i^2 - \frac{1}{2} \Lambda_i \frac{\partial \mathcal{F}^{\text{NS}}}{\partial \Lambda_i} = a_i^2 + \sum_{k=1}^\infty c_{i,k}(\mathbf{m}, \hbar) \mathbf{q}^k(\mathbf{a}) \,.
\end{equation}
Here, the coefficients $c_{i,k}$ are determined by the instanton calculus. Crucially, the identification of the two systems implies $u_i^{\text{BH}}(\omega) = u_i^{\text{gauge}}$.
The strategy for spectral determination therefore involves the inversion of the quantum mirror map. By numerically or analytically inverting Eq.~\eqref{eq:quantum_mirror_map}, one obtains the Coulomb moduli as functions of the physical frequency, $a_i = A_i(\mathbf{u}^{\text{BH}}(\omega))$. 
Substituting these frequency-dependent moduli back into the quantization condition Eq.~\eqref{eq:quantization_condition_vector} generates a closed system of algebraic equations:
\begin{equation}
    \label{eq:spectral_system}
    \left. \frac{\partial \mathcal{F}^{\text{NS}}}{\partial a_i} \right|_{a_j = A_j(\mathbf{u}^{\text{BH}}(\omega))} = 2\pi \hbar \left( n_i + \frac{1}{2} \right) \,.
\end{equation}
For a chosen set of mode numbers $\mathbf{n}$, the roots of this system yield the precise quasinormal frequencies $\omega$. Furthermore, this framework offers a promising perspective for higher-dimensional black hole backgrounds where the wave equation involves separation of variables. In such scenarios, the problem typically introduces unknown separation constants $\lambda_i$ alongside $\omega$. The multi-node structure of the quiver theory suggests that the additional quantization conditions (for $i > 1$) could, in principle, be utilized to constrain these separation constants. This implies that the full algebraic system has the capacity to simultaneously determine both the eigenfrequency and the separation constants, providing a systematic analytic alternative to traditional numerical relativity techniques.

\section{Conclusion}
\label{sec:conclusion}

In this paper, we have performed a systematic quantization of the SW geometry for $\mathcal{N}=2$, $\mathrm{SU}(2)^n$ linear quiver gauge theories. By applying the Weyl ordering prescription to the classical curve defined by the polynomial coefficients, we successfully derive the corresponding second-order differential equation. Through a transformation to the Schrödinger form and a detailed analysis of the singularity structure, we demonstrate that the quantum curve is isomorphic to the EHE with $n+3$ regular singular points.

Our primary mathematical result is the establishment of an explicit map between the polynomial coefficients of the SW curve and the canonical parameters of the EHE. By utilizing the relations established in the classical analysis, we are able to express the characteristic exponents and accessory parameters of the differential equation in terms of physical parameters. This generalizes the well-known correspondence for $N_f=4$ SQCD (the standard Heun-type case) to the entire family of $\mathrm{SU}(2)$ linear quivers.

While our focus has been on the formal derivation of the quantum geometry, this framework provides a powerful tool for analyzing physical problems in gravity. In particular, by matching the number of nodes in the quiver hierarchy with the spacetime dimension ($n=d-2$), our formalism enables the application of non-perturbative gauge theoretic techniques to the spectral problem of black holes with multiple regular singularities. Specifically, this isomorphism allows for the use of the NS free energy to determine the exact quantization conditions for QNMs, offering a robust analytic alternative to traditional numerical methods.

Looking forward, several directions for future research emerge. First, extending this analysis to linear quivers with higher rank gauge groups ($\mathrm{ SU(N)}$) would involve quantizing curves that lead to $N$-th order differential equations, a structure intimately related to $W_N$-algebra symmetry \cite{Wyllard:2009hg}. Secondly, incorporating irregular singularities into this hierarchy via confluence limits would extend our results to asymptotically free theories \cite{Piatek:2016kij}, corresponding to the Confluent Heun hierarchy and black hole backgrounds with different asymptotic behaviors. Finally, detailed calculations for black holes featuring more exotic singularity structures, building upon the potential applications outlined in Section \ref{sec:quantization_condition}, are reserved for future work.

\vspace{8pt}

$\\$
\noindent {\it Acknowledgements.} 
The authors thank  Xian-Hui Ge, Masataka Matsumoto, Yutaka Matsuo, Jean-Emile Bourgine, Futoshi Yagi, Yang Lei, Hongfei Shu and Rui-Dong Zhu
for helpful discussions.  K.Z. (Hong Zhang) is supported by a classified fund of Shanghai city.

\appendix

\section{Detailed Derivation of the Quantum Curve}
\label{app:derivation}

This appendix provides the algebraic details underlying the quantization of the SW curve and its subsequent reduction to the Schrödinger form, supplementing the discussion in Section \ref{sec:quantum_geometry}.

We begin with the classical algebraic curve defined by the polynomial equation:
\begin{equation}
    \sum_{k} f_k y^k x^2 - \sum_{k} c_k y^k x - \sum_{k} d_k y^k = 0 \,.
\end{equation}
The quantization is performed by promoting the coordinates to operators satisfying the commutation relation $[\hat{x}, \hat{y}] = \hbar \hat{y}$. In the coordinate representation where $\hat{y}$ acts multiplicatively and $\hat{x} = \hbar y \partial_y$, we adopt the Weyl symmetric ordering prescription $y^k x^m \to \hat{y}^k (\hat{x} + \frac{k\hbar}{2})^m$. The quantum spectral curve is then defined by the operator $\hat{\mathcal{D}}$ annihilating the wavefunction $\Psi(y)$:
\begin{equation}
    \left[ \sum_{k} f_k \hat{y}^k \left(\hat{x} + \frac{k\hbar}{2}\right)^2 - \sum_{k} c_k \hat{y}^k \left(\hat{x} + \frac{k\hbar}{2}\right) - \sum_{k} d_k \hat{y}^k \right] \Psi(y) = 0 \,.
    \label{eq:app_operator_def}
\end{equation}
To evaluate the action of the squared momentum operator, we expand the shift term explicitly:
\begin{equation}
    \left(\hbar y \partial_y + \frac{k\hbar}{2}\right)^2 = \hbar^2 y^2 \partial_y^2 + \hbar^2 (1+k) y \partial_y + \frac{k^2 \hbar^2}{4} \,.
\end{equation}
Substituting this expansion into \eqref{eq:app_operator_def} and collecting terms according to the order of the differential operator $\partial_y$, we obtain a second-order linear differential equation of the form:
\begin{equation}
    \left[ A(y) \frac{d^2}{dy^2} + B(y) \frac{d}{dy} + C(y) \right] \Psi(y) = 0 \,.
\end{equation}
The coefficient of the second derivative, $A(y)$, receives contributions exclusively from the quadratic term in $\hat{x}$:
\begin{equation}
    A(y) = \hbar^2 y^2 \sum_{k} f_k y^k \equiv \hbar^2 y^2 F_2(y) \,.
\end{equation}
The coefficient of the first derivative, $B(y)$, arises from both the linear and quadratic terms in $\hat{x}$. Collecting the relevant factors yields:
\begin{equation}
\begin{aligned}
    B(y) &= \hbar^2 \sum_{k} (1+k) f_k y^{k+1} - \hbar \sum_{k} c_k y^{k+1} \\
    &= \hbar^2 y \left( F_2(y) + y F_2'(y) \right) - \hbar y F_1(y) \,,
\end{aligned}
\end{equation}
where we have identified the polynomial $F_1(y) = \sum c_k y^k$ and used the identity $\sum k f_k y^k = y F_2'(y)$. Finally, the zero-th order term $C(y)$ contains the constant shifts and the $y$-dependent potential:
\begin{equation}
\begin{aligned}
    C(y) &= \frac{\hbar^2}{4} \sum_{k} k^2 f_k y^k - \frac{\hbar}{2} \sum_{k} k c_k y^k - \sum_{k} d_k y^k \\
    &= \frac{\hbar^2}{4} y (y F_2(y)')' - \frac{\hbar}{2} y F_1'(y) - F_0(y) \,,
\end{aligned}
\end{equation}
where $F_0(y) = \sum d_k y^k$. In deriving the last line, we utilized the operator identity for the scaling dimension, $\sum a_k k^2 y^k = (y\partial_y)^2 \sum a_k y^k$.

To facilitate the analysis of the quantum periods, we transform the operator equation into the standard Schrödinger form, $\psi''(y) + V(y)\psi(y) = 0$. This is achieved via the Liouville transformation $\Psi(y) = N(y)\psi(y)$, where the prefactor $N(y)$ is chosen to eliminate the first derivative term. The condition $2 A(y) N'(y) + B(y) N(y) = 0$ implies:
\begin{equation}
    \frac{N'(y)}{N(y)} = -\frac{B(y)}{2A(y)} = -\frac{1}{2} \left( \frac{1}{y} + \frac{F_2'(y)}{F_2(y)} - \frac{F_1(y)}{\hbar y F_2(y)} \right) \,.
\end{equation}
Integrating this relation yields the scaling factor:
\begin{equation}
    N(y) = \frac{1}{\sqrt{y F_2(y)}} \exp\left( \frac{1}{2\hbar} \int^y \frac{F_1(z)}{z F_2(z)} \, dz \right) \,.
\end{equation}
The resulting invariant potential $V(y)$ is determined by the standard relation:
\begin{equation}
    V(y) = \frac{C(y)}{A(y)} - \frac{1}{2} \left(\frac{B(y)}{A(y)}\right)' - \frac{1}{4} \left(\frac{B(y)}{A(y)}\right)^2 \,.
\end{equation}
Substituting the explicit expressions for $A(y)$, $B(y)$, and $C(y)$ derived above, and expanding the terms in powers of $\hbar$, we recover the exact potential given in Eq. \eqref{eq:potential_explicit}. This form highlights the quantum corrections to the classical geometry, with singular terms governed by the roots of $F_2(y)$ and $F_0(y)$.

\section{Singularity Structure and Potential Coefficients}
\label{app:residue_calc}

In this appendix, we derive the explicit expressions for the coefficients of the extended Heun potential \eqref{eq:EHE_canonical}. We calculate the local behavior of the invariant Schrödinger potential, defined in terms of the function $y(z)$ and its derivatives as:
\begin{equation}
    V(y) = \frac{C(y)}{A(y)} - \frac{1}{4}\left(\frac{B(y)}{A(y)}\right)^2 - \frac{1}{2} \frac{d}{dy}\left(\frac{B(y)}{A(y)}\right) \,,
    \label{eq:app_V_invariant}
\end{equation}
where the coefficient functions $A, B, C$ are given by
\begin{align}
    A(y) &= \hbar^2 y^2 F_2(y) \,, \nonumber \\
    B(y) &= \hbar^2 y \left( F_2(y) + y F_2'(y) \right) - \hbar y F_1(y) \,, \nonumber \\
    C(y) &= \frac{\hbar^2}{4} y (y F_2'(y))' - \frac{\hbar}{2} y F_1'(y) - F_0(y) \,.
\end{align}
We analyze the singularities of $V(y)$ at the finite roots $y_k$ of $F_2(y)$ and at the origin $y=0$. The strength of the double pole at a singularity $y=a$ is characterized by the coefficient $S_a = \lim_{y \to a} (y-a)^2 V(y)$.

Near a simple root $y_k$ of $F_2(y)$, the function $A(y)$ vanishes linearly, $A(y) \sim \hbar^2 y_k^2 F_2'(y_k)(y-y_k)$, while $B(y)$ remains finite. The singular behavior is dominated by the ratio $B/A$, which expands as
\begin{equation}
    \frac{B(y)}{A(y)} = \frac{\Lambda_k}{y-y_k} + \mathcal{O}(1) \,, \quad \text{with} \quad \Lambda_k \equiv 1 - \frac{F_1(y_k)}{\hbar y_k F_2'(y_k)} \,.
\end{equation}
Substituting this expansion into \eqref{eq:app_V_invariant}, we observe that the term $C/A$ scales as $(y-y_k)^{-1}$ and thus does not contribute to the quadratic pole. The contributions arise solely from the square and derivative of $B/A$. Summing the leading terms $-\frac{1}{4}\Lambda_k^2 (y-y_k)^{-2}$ and $\frac{1}{2}\Lambda_k (y-y_k)^{-2}$, we obtain the coefficient:
\begin{equation}
    S_{y_k} = \frac{1}{2}\Lambda_k - \frac{1}{4}\Lambda_k^2 \,.
\end{equation}

The analysis differs at the origin $y=0$ due to the $y^2$ prefactor in $A(y)$, which generates a double zero. Consequently, the term $C/A$ is singular enough to contribute to the quadratic pole. The asymptotic behaviors are given by $A(y) \sim \hbar^2 F_2(0) y^2$, $B(y) \sim \hbar [\hbar F_2(0) - F_1(0)] y$, and $C(y) \sim -F_0(0)$. Defining the parameter $\Lambda_0 = 1 - \frac{F_1(0)}{\hbar F_2(0)}$, the relevant expansion becomes:
\begin{equation}
    \frac{B(y)}{A(y)} \sim \frac{\Lambda_0}{y}\,, \quad \frac{C(y)}{A(y)} \sim -\frac{F_0(0)}{\hbar^2 F_2(0) y^2} \,.
\end{equation}
Collecting the contributions from all three terms in \eqref{eq:app_V_invariant} yields the coefficient at the origin:
\begin{equation}
    S_0 = -\frac{F_0(0)}{\hbar^2 F_2(0)} + \frac{1}{2}\Lambda_0 - \frac{1}{4}\Lambda_0^2 \,.
\end{equation}

Finally, the accessory parameters $K_k$ are determined by the residues of the simple poles at $y_k$. Expanding $V(y)$ to order $(y-y_k)^{-1}$, we identify the residue as:
\begin{equation}
    \frac{K_k}{y_k} \equiv \text{Res}_{y=y_k} V(y) = \lim_{y \to y_k} \frac{d}{dy}\left[ (y-y_k)^2 V(y) \right] \,.
\end{equation}
In terms of the coefficient functions, this is evaluated as:
\begin{equation}
    \text{Res}_{y=y_k} V(y) = \frac{C(y_k)}{A'(y_k)} - \frac{1}{2} \frac{B(y_k)}{A'(y_k)} \left( \frac{B'(y_k)}{A'(y_k)} - \frac{B(y_k)A''(y_k)}{2[A'(y_k)]^2} \right) \,,
\end{equation}
where the derivatives are evaluated at the root $y_k$. This completes the derivation of the potential parameters.

\section{Derivation of the Polynomial Numerator}
\label{app:polynomial_derivation}

In this appendix, we analyze the polynomial numerator $\mathcal{P}(y)$ determining the quantum potential $V(y) = \frac{\mathcal{P}(y)}{4\hbar^2 y^2 F_2(y)^2}$. Our goal is to derive the recurrence relations for the expansion coefficients $\mathcal{A}_m$, defined by $\mathcal{P}(y) = \sum_{m} \mathcal{A}_m y^m$. We rely on the $\hbar$-expansion of the coefficient functions $A(y)$, $B(y)$, and $C(y)$ detailed in Appendix \ref{app:derivation}.

The numerator is given explicitly by:
\begin{equation}
    \mathcal{P}(y) = \frac{1}{\hbar^2 y^2} \left[ 4AC - B^2 - 2AB' + 2BA' \right] \,.
\end{equation}
We first verify the consistency of the quantization by ensuring that terms of odd order in $\hbar$ vanish. Using the components $A_2, B_1, B_2$, and $C_1$ identified in the previous section, the $\mathcal{O}(\hbar^3)$ contribution to the bracket is:
\begin{equation}
    \mathcal{T}_{\hbar^3} = 4 A_2 C_1 - 2 B_1 B_2 - 2(A_2 B_1' - B_1 A_2') \,.
\end{equation}
Substituting the explicit forms (e.g., $C_1 = -\frac{1}{2}y F_1'$), we find:
\begin{align}
    \mathcal{T}_{\hbar^3} &= -2 y^3 F_2 F_1' - 2(-y F_1)(y F_2 + y^2 F_2') \nonumber \\
    &\quad\, - 2(y^2 F_2)(-F_1 - y F_1') + 2(-y F_1)(2y F_2 + y^2 F_2') \nonumber \\
    &= y^3 F_2 F_1' (-2 + 2) + y^3 F_1 F_2' (2 - 2) \nonumber \\
    &\quad\, + y^2 F_1 F_2 (2 + 2 - 4) \equiv 0 \,.
\end{align}
This cancellation guarantees that the potential admits an expansion in even powers of $\hbar$.

The non-vanishing contributions arise from the $\mathcal{O}(\hbar^0)$ and $\mathcal{O}(\hbar^2)$ terms. The classical limit yields:
\begin{equation}
    \mathcal{P}_{\text{cl}}(y) = - (F_1^2 + 4 F_0 F_2) \,.
\end{equation}
The quantum correction terms, corresponding to $\mathcal{O}(\hbar^4)$ in the bracket (divided by the $\hbar^2 y^2$ prefactor), simplify to:
\begin{equation}
    \mathcal{P}_{\text{quant}}(y) = \hbar^2 \left[ F_2^2 - y F_2 F_2' + y^2 (F_2')^2 - y^2 F_2 F_2'' \right] \,.
    \label{eq:P_quant_def}
\end{equation}

To obtain the series coefficients $\mathcal{A}_m$, we insert the polynomial expansions $F_2(y) = \sum f_j y^j$, $F_1(y) = \sum c_j y^j$, and $F_0(y) = \sum d_j y^j$.
For the quantum part $\mathcal{P}_{\text{quant}}$, we consider the action of the differential operators in Eq.~\eqref{eq:P_quant_def} on the product of series. A generic term $f_j f_k y^{j+k}$ transforms as:
\begin{equation}
    y^{j+k} f_j f_k \left[ 1 - k + jk - k(k-1) \right] = y^{j+k} f_j f_k (1 + jk - k^2) \,.
\end{equation}
We are interested in the coefficient of a specific power $y^m$ (where $j+k=m$). Since the summation runs over all $j$ and $k$, we symmetrize the weight factor with respect to the interchange $j \leftrightarrow k$:
\begin{equation}
    W_{j,k}^{\text{sym}} = \frac{1}{2} \left[ (1 + jk - k^2) + (1 + kj - j^2) \right] = 1 + jk - \frac{1}{2}(j^2 + k^2) \,.
\end{equation}
Using the identity $(j-k)^2 = j^2 + k^2 - 2jk$, this weight simplifies to:
\begin{equation}
    W_{j,k}^{\text{sym}} = 1 - \frac{1}{2}(j-k)^2 \,.
\end{equation}
Combining this with the classical contribution $\mathcal{P}_{\text{cl}}$, which involves the coefficients $c_j$ and $d_j$, we arrive at the final formula for the coefficients $\mathcal{A}_m$:
\begin{equation}
    \label{eq:Ak_coefficients2}
    \mathcal{A}_m = \hbar^2 \sum_{j=0}^{m} f_j f_{m-j} \left[ 1 - \frac{1}{2}(2j - m)^2 \right] - \sum_{j=0}^{m} \left( c_j c_{m-j} + 4 d_j f_{m-j} \right) \,.
\end{equation}
This completes the derivation of the potential parameters.

\bibliographystyle{JHEP}
\bibliography{YWZ}

\providecommand{\href}[2]{#2}\begingroup\raggedright\begin{thebibliography}{10}

\bibitem{Seiberg:1994rs}
N.~Seiberg and E.~Witten, \emph{{Electric - magnetic duality, monopole
  condensation, and confinement in N=2 supersymmetric Yang-Mills theory}},
  \href{http://dx.doi.org/10.1016/0550-3213(94)90124-4}{\emph{Nucl. Phys. B}
  {\bf 426} (1994) 19--52}, [\href{http://arxiv.org/abs/hep-th/9407087}{{\tt
  hep-th/9407087}}].

\bibitem{Seiberg:1994aj}
N.~Seiberg and E.~Witten, \emph{{Monopoles, duality and chiral symmetry
  breaking in N=2 supersymmetric QCD}},
  \href{http://dx.doi.org/10.1016/0550-3213(94)90214-3}{\emph{Nucl. Phys. B}
  {\bf 431} (1994) 484--550}, [\href{http://arxiv.org/abs/hep-th/9408099}{{\tt
  hep-th/9408099}}].

\bibitem{Aminov:2020yma}
G.~Aminov, A.~Grassi and Y.~Hatsuda, \emph{{Black Hole Quasinormal Modes and
  Seiberg\textendash{}Witten Theory}},
  \href{http://dx.doi.org/10.1007/s00023-021-01137-x}{\emph{Annales Henri
  Poincare} {\bf 23} (2022) 1951--1977},
  [\href{http://arxiv.org/abs/2006.06111}{{\tt 2006.06111}}].

\bibitem{Aminov:2023jve}
G.~Aminov, P.~Arnaudo, G.~Bonelli, A.~Grassi and A.~Tanzini, \emph{{Black hole
  perturbation theory and multiple polylogarithms}},
  \href{http://arxiv.org/abs/2307.10141}{{\tt 2307.10141}}.

\bibitem{Nekrasov:2009rc}
N.~A. Nekrasov and S.~L. Shatashvili, \emph{{Quantization of Integrable Systems
  and Four Dimensional Gauge Theories}},  in \emph{{16th International Congress
  on Mathematical Physics}}, pp.~265--289, 8, 2009.
\newblock \href{http://arxiv.org/abs/0908.4052}{{\tt 0908.4052}}.
\newblock \href{http://dx.doi.org/10.1142/9789814304634_0015}{DOI}.

\bibitem{Gaiotto:2009we}
D.~Gaiotto, \emph{{N=2 dualities}},
  \href{http://dx.doi.org/10.1007/JHEP08(2012)034}{\emph{JHEP} {\bf 08} (2012)
  034}, [\href{http://arxiv.org/abs/0904.2715}{{\tt 0904.2715}}].

\bibitem{Witten:1997sc}
E.~Witten, \emph{{Solutions of four-dimensional field theories via M-theory}},
  \href{http://dx.doi.org/10.1201/9781482268737-38}{\emph{Nucl. Phys. B} {\bf
  500} (1997) 3--42}, [\href{http://arxiv.org/abs/hep-th/9703166}{{\tt
  hep-th/9703166}}].

\bibitem{Alday:2009aq}
L.~F. Alday, D.~Gaiotto and Y.~Tachikawa, \emph{{Liouville Correlation
  Functions from Four-dimensional Gauge Theories}},
  \href{http://dx.doi.org/10.1007/s11005-010-0369-5}{\emph{Lett. Math. Phys.}
  {\bf 91} (2010) 167--197}, [\href{http://arxiv.org/abs/0906.3219}{{\tt
  0906.3219}}].

\bibitem{Bianchi:2021mft}
M.~Bianchi, D.~Consoli, A.~Grillo and J.~F. Morales, \emph{{More on the SW-QNM
  correspondence}},
  \href{http://dx.doi.org/10.1007/JHEP01(2022)024}{\emph{JHEP} {\bf 01} (2022)
  024}, [\href{http://arxiv.org/abs/2109.09804}{{\tt 2109.09804}}].

\bibitem{Consoli:2022eey}
D.~Consoli, F.~Fucito, J.~F. Morales and R.~Poghossian, \emph{{CFT description
  of BH\textquoteright{}s and ECO\textquoteright{}s: QNMs, superradiance,
  echoes and tidal responses}},
  \href{http://dx.doi.org/10.1007/JHEP12(2022)115}{\emph{JHEP} {\bf 12} (2022)
  115}, [\href{http://arxiv.org/abs/2206.09437}{{\tt 2206.09437}}].

\bibitem{Gorsky:2015toa}
A.~Gorsky, A.~Milekhin and N.~Sopenko, \emph{{The Condensate from Torus
  Knots}}, \href{http://dx.doi.org/10.1007/JHEP09(2015)102}{\emph{JHEP} {\bf
  09} (2015) 102}, [\href{http://arxiv.org/abs/1506.06695}{{\tt 1506.06695}}].

\bibitem{CarneirodaCunha:2015qln}
B.~Carneiro~da Cunha and F.~Novaes, \emph{{Kerr{\textendash}de Sitter greybody
  factors via isomonodromy}},
  \href{http://dx.doi.org/10.1103/PhysRevD.93.024045}{\emph{Phys. Rev. D} {\bf
  93} (2016) 024045}, [\href{http://arxiv.org/abs/1508.04046}{{\tt
  1508.04046}}].

\bibitem{Kokkotas:1999bd}
K.~D. Kokkotas and B.~G. Schmidt, \emph{{Quasinormal modes of stars and black
  holes}}, \href{http://dx.doi.org/10.12942/lrr-1999-2}{\emph{Living Rev. Rel.}
  {\bf 2} (1999) 2}, [\href{http://arxiv.org/abs/gr-qc/9909058}{{\tt
  gr-qc/9909058}}].

\bibitem{Berti:2009kk}
E.~Berti, V.~Cardoso and A.~O. Starinets, \emph{{Quasinormal modes of black
  holes and black branes}},
  \href{http://dx.doi.org/10.1088/0264-9381/26/16/163001}{\emph{Class. Quant.
  Grav.} {\bf 26} (2009) 163001}, [\href{http://arxiv.org/abs/0905.2975}{{\tt
  0905.2975}}].

\bibitem{Konoplya:2011qq}
R.~A. Konoplya and A.~Zhidenko, \emph{{Quasinormal modes of black holes: From
  astrophysics to string theory}},
  \href{http://dx.doi.org/10.1103/RevModPhys.83.793}{\emph{Rev. Mod. Phys.}
  {\bf 83} (2011) 793--836}, [\href{http://arxiv.org/abs/1102.4014}{{\tt
  1102.4014}}].

\bibitem{Horowitz:1999jd}
G.~T. Horowitz and V.~E. Hubeny, \emph{{Quasinormal modes of AdS black holes
  and the approach to thermal equilibrium}},
  \href{http://dx.doi.org/10.1103/PhysRevD.62.024027}{\emph{Phys. Rev. D} {\bf
  62} (2000) 024027}, [\href{http://arxiv.org/abs/hep-th/9909056}{{\tt
  hep-th/9909056}}].

\bibitem{Nekrasov:2002qd}
N.~A. Nekrasov, \emph{{Seiberg-Witten prepotential from instanton counting}},
  \href{http://dx.doi.org/10.4310/ATMP.2003.v7.n5.a4}{\emph{Adv. Theor. Math.
  Phys.} {\bf 7} (2003) 831--864},
  [\href{http://arxiv.org/abs/hep-th/0206161}{{\tt hep-th/0206161}}].

\bibitem{Gukov:2011qp}
S.~Gukov and P.~Sulkowski, \emph{{A-polynomial, B-model, and Quantization}},
  \href{http://dx.doi.org/10.1007/JHEP02(2012)070}{\emph{JHEP} {\bf 02} (2012)
  070}, [\href{http://arxiv.org/abs/1108.0002}{{\tt 1108.0002}}].

\bibitem{Mironov:2009uv}
A.~Mironov and A.~Morozov, \emph{{Nekrasov Functions and Exact Bohr-Zommerfeld
  Integrals}}, \href{http://dx.doi.org/10.1007/JHEP04(2010)040}{\emph{JHEP}
  {\bf 04} (2010) 040}, [\href{http://arxiv.org/abs/0910.5670}{{\tt
  0910.5670}}].

\bibitem{Batic:2007it}
D.~Batic and H.~Schmid, \emph{{Heun equation, Teukolsky equation, and type-D
  metrics}}, \href{http://dx.doi.org/10.1063/1.2720277}{\emph{J. Math. Phys.}
  {\bf 48} (2007) 042502}, [\href{http://arxiv.org/abs/gr-qc/0701064}{{\tt
  gr-qc/0701064}}].

\bibitem{Staicova:2016awo}
D.~Staicova and P.~Fiziev, \emph{{The Heun functions and their applications in
  astrophysics}},
  \href{http://dx.doi.org/10.1007/978-981-10-2636-2_20}{\emph{Springer Proc.
  Math. Stat.} {\bf 191} (2016) 303--308},
  [\href{http://arxiv.org/abs/1601.04021}{{\tt 1601.04021}}].

\bibitem{Lei:2023mqx}
Y.~Lei, H.~Shu, K.~Zhang and R.-D. Zhu, \emph{{Quasinormal modes of C-metric
  from SCFTs}}, \href{http://dx.doi.org/10.1007/JHEP02(2024)140}{\emph{JHEP}
  {\bf 02} (2024) 140}, [\href{http://arxiv.org/abs/2308.16677}{{\tt
  2308.16677}}].

\bibitem{Ge:2024jdx}
X.-H. Ge, M.~Matsumoto and K.~Zhang, \emph{{Duality between Seiberg-Witten
  theory and black hole superradiance}},
  \href{http://dx.doi.org/10.1007/JHEP05(2024)336}{\emph{JHEP} {\bf 05} (2024)
  336}, [\href{http://arxiv.org/abs/2402.17441}{{\tt 2402.17441}}].

\bibitem{Ge:2025yqk}
X.-H. Ge, M.~Matsumoto and K.~Zhang, \emph{{Massive vector field perturbations
  in the Schwarzschild spacetime from supersymmetric gauge theory}},
  \href{http://dx.doi.org/10.1103/3dwl-gg8g}{\emph{Phys. Rev. D} {\bf 112}
  (2025) 046024}, [\href{http://arxiv.org/abs/2502.15627}{{\tt 2502.15627}}].

\bibitem{Kodama:2003jz}
H.~Kodama and A.~Ishibashi, \emph{{A Master equation for gravitational
  perturbations of maximally symmetric black holes in higher dimensions}},
  \href{http://dx.doi.org/10.1143/PTP.110.701}{\emph{Prog. Theor. Phys.} {\bf
  110} (2003) 701--722}, [\href{http://arxiv.org/abs/hep-th/0305147}{{\tt
  hep-th/0305147}}].

\bibitem{Slavyanov:2000sf}
S.~Y. Slavyanov and W.~Lay, \emph{Special Functions: A Unified Theory Based on
  Singularities}.
\newblock Oxford Mathematical Monographs. Oxford University Press, Oxford,
  2000.

\bibitem{Imaizumi:2022dgj}
K.~Imaizumi, \emph{{Exact conditions for quasi-normal modes of extremal
  M5-branes and exact WKB analysis}},
  \href{http://dx.doi.org/10.1016/j.nuclphysb.2023.116221}{\emph{Nucl. Phys. B}
  {\bf 992} (2023) 116221}, [\href{http://arxiv.org/abs/2212.04738}{{\tt
  2212.04738}}].

\bibitem{Hatsuda:2019eoj}
Y.~Hatsuda, \emph{{Quasinormal modes of black holes and Borel summation}},
  \href{http://dx.doi.org/10.1103/PhysRevD.101.024008}{\emph{Phys. Rev. D} {\bf
  101} (2020) 024008}, [\href{http://arxiv.org/abs/1906.07232}{{\tt
  1906.07232}}].

\bibitem{Aganagic:2011mi}
M.~Aganagic, M.~C.~N. Cheng, R.~Dijkgraaf, D.~Krefl and C.~Vafa, \emph{{Quantum
  Geometry of Refined Topological Strings}},
  \href{http://dx.doi.org/10.1007/JHEP11(2012)019}{\emph{JHEP} {\bf 11} (2012)
  019}, [\href{http://arxiv.org/abs/1105.0630}{{\tt 1105.0630}}].

\bibitem{Grassi:2019coc}
A.~Grassi, J.~Gu and M.~Mari\~no, \emph{{Non-perturbative approaches to the
  quantum Seiberg-Witten curve}},
  \href{http://dx.doi.org/10.1007/JHEP07(2020)106}{\emph{JHEP} {\bf 07} (2020)
  106}, [\href{http://arxiv.org/abs/1908.07065}{{\tt 1908.07065}}].

\bibitem{Matone:1995rx}
M.~Matone, \emph{{Instantons and recursion relations in N=2 SUSY gauge
  theory}}, \href{http://dx.doi.org/10.1016/0370-2693(95)00920-G}{\emph{Phys.
  Lett. B} {\bf 357} (1995) 342--348},
  [\href{http://arxiv.org/abs/hep-th/9506102}{{\tt hep-th/9506102}}].

\bibitem{Flume:2004rp}
R.~Flume, F.~Fucito, J.~F. Morales and R.~Poghossian, \emph{{Matone's relation
  in the presence of gravitational couplings}},
  \href{http://dx.doi.org/10.1088/1126-6708/2004/04/008}{\emph{JHEP} {\bf 04}
  (2004) 008}, [\href{http://arxiv.org/abs/hep-th/0403057}{{\tt
  hep-th/0403057}}].

\bibitem{Wyllard:2009hg}
N.~Wyllard, \emph{{A(N-1) conformal Toda field theory correlation functions
  from conformal N = 2 SU(N) quiver gauge theories}},
  \href{http://dx.doi.org/10.1088/1126-6708/2009/11/002}{\emph{JHEP} {\bf 11}
  (2009) 002}, [\href{http://arxiv.org/abs/0907.2189}{{\tt 0907.2189}}].

\bibitem{Piatek:2016kij}
M.~R. Piatek and A.~R. Pietrykowski, \emph{{Irregular blocks, $\mathcal{N} = 2$
  gauge theory and Mathieu system}},
  \href{http://dx.doi.org/10.1088/1742-6596/670/1/012041}{\emph{J. Phys. Conf.
  Ser.} {\bf 670} (2016) 012041}.

\end{thebibliography}\endgroup

\end{document}